\documentclass[superscriptaddress,twocolumn,prb,
               showpacs,preprintnumbers,floatfix]{revtex4}
\usepackage{amsmath,amssymb}
\usepackage{epsfig,graphics,color,calc}

\usepackage{color}
\usepackage{graphicx}
\usepackage{mathrsfs}

\newcommand{\rr}[1]{{\normalfont\textrm{#1}}}

\newcommand{\bb}[1]{{\mathbb{#1}}}

\definecolor{light}{gray}{.9}

\begin{document}

\title{Phase coexistence in consolidating porous media}

\author{Emilio N.M.\ Cirillo}
\email{cirillo@dmmm.uniroma1.it}
\affiliation{Dipartimento Me.\ Mo.\ Mat., 
Universit\`a degli Studi di Roma ``La Sapienza," 
via A.\ Scarpa 16, 00161 Roma, Italy}

\author{Nicoletta Ianiro}
\email{ianiro@dmmm.uniroma1.it}
\affiliation{Dipartimento Me.\ Mo.\ Mat., 
Universit\`a degli Studi di Roma ``La Sapienza," 
via A.\ Scarpa 16, 00161 Roma, Italy}

\author{Giulio Sciarra}
\email{giulio.sciarra@uniroma1.it}
\affiliation{Dipartimento di Ingegneria Chimica Materiali Ambiente,
Universit\`a degli Studi di Roma ``La Sapienza," 
via Eudossiana 18, 00184 Roma, Italy}


\begin{abstract}
The appearence of the fluid--rich phase in saturated porous media 
under the effect of an external pressure is investigated. 
For this purpose we introduce a two field second gradient model 
allowing the complete description of the phenomenon.
We study the coexistence profile between poor and rich fluid phases and 
we show that for a suitable choice of the parameters non--monotonic 
interfaces show up at coexistence.
\end{abstract}




\maketitle

\section{Introduction}
\label{s:introduzione}
\par\noindent
When a porous medium is plunged in an infinite fluid reservoir, 
the solid matrix absorbs fluid until an equilibrium 
state is reached. 
Many interesting features of this swelling phenomenon have been
demonstrated experimentally. 
The amount of swelled fluid can be controlled 
via different external parameters such as 
the fluid pressure in the reservoir~\cite{GGRS,CC02}, that is via its chemical 
potential, 
the fluid velocity \cite{Nichols94},
or a mechanical pressure exerted on the solid 
\cite{Vardoulakis04_1,CIS01,CIS02}.

The solid--fluid segregation in consolidation 
is seen when, depending on the external pressure acting on the 
porous material, phases differing in fluid content are observed. 
This problem has been addressed\cite{CIS01} by the authors in the framework 
of continuum mechanics adopting in particular a first gradient model;
the existence of two different phases 
depending on the external pressure has been proved. In that context the 
interesting question of the coexistence of the two phases could not be posed 
for the first gradient nature of the model. 
In this paper we propose a 
more general two field 
one dimensional
second gradient model to study the profiles connecting two 
coexisting phases
and the formation of critical droplets, if any, of one phase 
into the other.

The adopted approach in modeling the behavior of porous continua 
is essentially based on a pure solid Lagrangian description of motion, 
referring kinematics to the reference configuration of the porous skeleton 
(see Section~\ref{s:setup}). 

The constitutive model is purely phenomenological, which means that the 
overall potential energy, regarded as a function of the strain of the skeleton 
and the fluid mass density (per unit volume, in the solid reference 
configuration), is built up in such a way to describe the existence of two 
states of equilibrium: the solid--rich and the fluid--rich phase. 
Thus no refined description of solid grain connectivity, as well as 
connection among regions with different porosity is available in such a model. 
Conversely the constitutive state parameters are selected so as  
to describe the showing up of the fluid--rich phase, which is 
possibly associated to the occurrence of fluid segregation.

\section{The poromechanics setup}
\label{s:setup}
\par\noindent
Let $B_\rr{s}\subset\bb{R}$ be the \textit{reference}
configurations
for the solid and fluid components~\cite{Coussy}. 
The \textit{solid placement} is a $C^2$--diffeomorphism
$\chi_\rr{s}(\cdot,t):B_\rr{s}\to\bb{R}$ such that 
$\chi_\rr{s}(X_\rr{s},t)$ is the 
position occupied at time $t$ by the solid particle 
$X_\rr{s}$ in the reference configuration $B_\rr{s}$.
Consider~\cite{SIIM} 
$\phi(\cdot,t):B_\rr{s}\to\bb{R}$ 
such that 
$\phi(X_\rr{s},t)$ is 
the fluid particle which at time $t$ occupies the 
same position of the solid particle $X_\rr{s}$.
Assume also $\phi(\cdot,t)$ to be a $C^2$--diffeomorphism,
thus  
the map $\phi(\cdot,t)$ associate univocally a solid particle 
to a fluid one and vice versa.
The \textit{fluid placement} map 
$\chi_\rr{f}(\cdot,t):\bb{R}\to\bb{R}$,
giving the position of a fluid particle $X_\rr{f}$,
is defined as 
$\chi_\rr{f}(X_\rr{f},t):=\chi_\rr{s}(\phi^{-1}(X_\rr{f},t),t)$.
The \textit{current configuration} $\chi_\rr{s}(B_\rr{s},t)$ at time 
$t$ is the set of positions 
of the superposed solid and fluid particles.

Let 
$J_\rr{s}(X_\rr{s},t):=|\partial\chi_\rr{s}(X_\rr{s},t)/\partial X_\rr{s}|$
be the Jacobian of the placement map 
$\chi_\rr{s}(\cdot,t)$ measuring the ratio between current and 
reference volumes of the solid component; we let 
$\varepsilon(X_\rr{s},t):=(J_\rr{s}(X_\rr{s},t)^2-1)/2$ 
be the \textit{strain field}.
Let $\varrho_{0,\alpha}(X_\alpha)$ with $\alpha=\rr{s},\rr{f}$,
be the solid and fluid reference \textit{densities}; we define the 
\textit{fluid mass density} field
$m(X_\rr{s},t)
 :=\varrho_{0,\rr{f}}(\phi(X_\rr{s},t))
   \partial\phi(X_\rr{s},t)/\partial X_\rr{s}$.
Assuming that the mass is conserved, it is not difficult to prove~\cite{SIIM}
that the field $m$ can be interpreted as 
the fluid mass density measured w.r.t.\ (with respect to) the 
solid reference volume.

Assume, now, that the \textit{Lagrangian density} 
$\mathscr{L}
(\dot\chi_\rr{s},\dot\phi,
             \chi''_\rr{s},\phi'',\chi'_\rr{s},\phi',\chi_\rr{s},\phi)$
of the 
system is in the form 
\begin{equation}
\label{lag}
\mathscr{L}
 =T(\dot\chi_\rr{s},\dot\phi,\chi_\rr{s},\phi)
  -\Phi(\chi''_\rr{s},\phi'',\chi'_\rr{s},\phi',\chi_\rr{s},\phi)
\end{equation}
where $T$ is the \textit{kinetic energy density} and $\Phi$
is the \textit{overall potential energy density} accounting for both 
the internal and the external forces.
In~(\ref{lag}) we have denoted with the dot 
the derivative taken w.r.t.\ time and with the prime the derivative 
w.r.t.\ the solid reference space variable.
The equation of motion for the two fields $\chi_\rr{s}$ and $\phi$ can be 
derived assuming that the possible motions of the system in an 
interval of time $(t_1,t_2)\subset\bb{R}$
are those such that the fields $\chi_\rr{s}$ and $\phi$ are 
extremals
for the \textit{action functional}
\begin{equation}
\label{azione}
 A(\dot\chi_\rr{s},\dots,\phi)
 :=
 \int_{B_\rr{s}}\rr{d}X_\rr{s}\int_{t_1}^{t_2}\rr{d}t\,
 \mathscr{L}(
            \dot\chi_\rr{s},
            \dots,
            \phi)
\end{equation}
in correspondence of the independent variations 
of the two fields
$\chi_\rr{s}$ and $\phi$ on $B_\rr{s}\times(t_1,t_2)$.
In other words any possible motion of the system in the 
considered interval is a solution of the Euler--Lagrange equations
associated to the variational principle $\delta A=0$.

If one is interested to find equilibrium profiles $\chi_\rr{s}(X_\rr{s})$ 
and $\phi(X_\rr{s})$ of the system, 
namely, the solutions of the equations of motion independent of time, 
since the kinetic energy associated to those profiles is equal to zero,
the Lagrangian density reduces to minus the potential energy density.
In this case 
the action is given by (\ref{azione}) where the time integral gives a not 
essential multiplicative constant, 
and the variational principle associated to such an action gives the 
seeked for equilibrium profiles.

Assume, now, that the effect of the internal forces exchanged by the 
solid and fluid particles and that of the conservative external 
fields can be described via a potential energy density
$\Phi(m',\varepsilon',m,\varepsilon)$ depending on the 
kinematic fields $\chi_\rr{s}$ and $\phi$ only through the 
strain and the fluid mass density fields.
Note that 
the strain depends only on $\chi_\rr{s}$ and 
the fluid mass density only on $\phi$, hence the independent variations
of those primitive fields reflect on independent variations of $\varepsilon$
and $m$. 
Thus, limiting the study to boundary value problems 
expressed in terms of the fields $\varepsilon$ and $m$, 
we can treat the fields $\varepsilon$ and $m$ as primitive, 
consider their independent variations 
and look for the equilibrium profiles 
$\varepsilon(X_\rr{s})$ and $m(X_\rr{s})$ starting from the variational 
principle 
\begin{equation}
\label{variazionale03}
\delta\int_{B_\rr{s}}\rr{d}X_\rr{s}\,
\Phi(m'(X_\rr{s}),\varepsilon'(X_\rr{s}),m(X_\rr{s}),\varepsilon(X_\rr{s}))=0
\end{equation}

Finally, we can derive, starting from (\ref{variazionale03}), the equations
governing the equilibrium profiles $\varepsilon$ and $m$. By 
computing the variation of the action functional on 
$B_\rr{s}=(\ell_1,\ell_2)$, with $\ell_1,\ell_2\in\bb{R}$, we get the 
Euler--Lagrange equations
\begin{equation}
\label{moto06}
       \frac{\partial\Phi}{\partial\varepsilon}
       -\frac{\rr{d}}{\rr{d}X_\rr{s}}
        \frac{\partial\Phi}{\partial\varepsilon'}
       =0
       \;\;\;\textrm{ and }\;\;\;
        \frac{\partial\Phi}{\partial m}
       -\frac{\rr{d}}{\rr{d}X_\rr{s}}
       \frac{\partial\Phi}{\partial m'}
       =0
\end{equation}
with boundary conditions ensuring that 
\begin{equation}
\label{moto07}
  \Big[
       \frac{\partial\Phi}{\partial\varepsilon'}\delta\varepsilon
       +
       \frac{\partial\Phi}{\partial m'}\delta m
  \Big]_{\ell_1}^{\ell_2}
=0
\end{equation}
where $\delta\varepsilon$ and $\delta m$ are, respectively, the 
variations of the strain and fluid mass density fields.
For instance Dirichelet boundary conditions would do the job, since 
we would have $\delta\varepsilon(\ell_1)=\delta\varepsilon(\ell_2)=0$ and 
$\delta m(\ell_1)=\delta m(\ell_2)=0$.
But for potential energy densities $\Phi$ at least 
quadratic in the derivatives 
$\varepsilon'$ and $m'$ even Neumann boundary conditions would be 
acceptable.
In the sequel we shall refer to any solution of the 
Euler--Lagrange equations (\ref{moto06}) with suitable 
boundary conditions as an \textit{equilibrium profile} of 
the system corresponding to the chosen boundary conditions.

The model (\ref{variazionale03})
is called a \textit{first gradient} model 
if the potential energy density $\Phi$
does not depend on the first 
derivatives of the strain and of the liquid density, 
otherwise the model is said a \textit{second gradient} model. 
This way of classifying the models
is related to the fact that both $\varepsilon$ and $m$ depends, by definition,
on the gradient of the 
primitive kinetic fields $\chi_\rr{s}$ and $\phi$.

Second gradient theories are suitable to be developed for modeling
stress/strain concentration due, for instance, to the presence
of geometrical singularities (crack propagation in fracture 
mechanics\cite{Unger}) 
or
phase transitions as in the case of wetting\cite{Seppecher,deGennes}.
In particular
second gradient poromechanics has been recently 
formulated~\cite{Sciarra,SIIM} 
extending the standard arguments of the Biot theory~\cite{biot01}.
Such a model addresses the description of those deformation phenomena 
which occur at the same length scale as that where high gradients 
in deformation can be detected. 
Classical poromechanics\cite{Coussy} is not able to describe these phenomena: 
the macroscopic model is regarded in that case as the average of a 
microscopic one where a kind of stationarity assumption\cite{torquato} 
(spatial ergodicity) on the random field which characterizes the 
microscopic mechanical properties of the material has been formulated. 
This allows for replacing ensemble averages with volume averages 
insofar as the characteristic size of the heterogeneities is much smaller 
than the  typical length scale of the reference volume element (RVE). 
If this is no more the case, the classical assumptions of uniform strain 
(stress) or periodic boundary conditions, for every reference volume, 
are no more valid, but, conversely, macroscopic strain gradient plays a 
crucial role in specifying the state of stress/strain inside the RVE itself.

The goal, here, is to formulate a second gradient poromechanical 
model for describing the transition from the standard Biot--like 
equilibrium, associated to a compacted solid, versus the 
fluid--segregated phase describing duct thinnering in the matrix
and, consequently, fluid mass concentration in the pores.
Thus, 
we consider a model with total potential energy density  
in the form 
\begin{equation}
\label{modello}
\Phi(m',\varepsilon',m,\varepsilon)
=
K(m',\varepsilon',m,\varepsilon)
+
\Psi(m,\varepsilon)
\end{equation}
where
$K$ is a polynomial quadratic function of $\varepsilon'$ and $m'$,
$\Psi$ is a differentiable function positively diverging along any radial 
direction in the plane $\varepsilon$--$m$,
having at least a local minimum,
and whose stationary points are isolated.
Since $K$ is quadratic, we have that a constant solution 
of the Euler--Lagrange problem (\ref{moto06}) must necessarily 
satisfy the equations $\Psi_\varepsilon=0$ and $\Psi_m=0$; in other 
words a constant profile must be constantly equal to an extremal 
point of the first gradient part $\Psi$ of the total potential energy.

We then let
a \textit{phase} of the model to be a constant equilibrium profile 
equal to one of the local minima of the function $\Psi$.

Note that the Euler--Lagrange problem 
(\ref{moto06}) and (\ref{moto07}) 
for the first gradient model associated to (\ref{modello}), 
namely the one obtained for $K=0$,
is the system of algebraic equations 
$\Psi_m=0$ and $\Psi_\varepsilon=0$. Since the stationary points of the 
two variable function $\Psi$ are isolated, we have that  
the equilibrium profiles for such a model are necessarily constant 
functions of $X_\rr{s}\in B_\rr{s}$ equal to  
the values of the 
stationary points of $\Psi$.

Hence, in the case of a first gradient model it is not possible 
to discuss phase coexistence, since there exist only continuous 
constant equilibrium profiles.
On the other hand, 
in second gradient models, different (not constant) equilibrium profiles 
can exist. This fact allows us to pose the problem of the 
coexistence of two existing phases.
Suppose that the model exhibits the two phases 
$(m_1,\varepsilon_1)$
and
$(m_2,\varepsilon_2)$;
a \textit{connection}~\cite{AF} between those phases is 
an equilibrium profile  
$m,\varepsilon$ of the action functional 
on $B_\rr{s}=(-\infty,+\infty)$
satisfying the boundary conditions 
$m(-\infty)=m_1$,
$m(+\infty)=m_2$,
$\varepsilon(-\infty)=\varepsilon_1$,
and
$\varepsilon(+\infty)=\varepsilon_2$.
We say that the two considered phases \textit{coexist} if and only if a 
connection does exist.


\section{The model}
\label{s:ilmodello}
\par\noindent
We study, now, a particular poroelastic model and in that framework we 
discuss the existence of the consolidation phase transition and prove 
the coexistence of the two phases for a particular value of the external 
pressure.
More precisely we consider the poroelastic system with overall potential 
energy density (\ref{modello}) with
\begin{equation}
\label{secondo000}
 K(m',\varepsilon')
:=
 \frac{1}{2}[k_1(\varepsilon')^2+2k_2\varepsilon' m'+k_3(m')^2]
\end{equation}
with $k_1,k_3>0$, $k_2\in\bb{R}$ such that $k_1k_3-k_2^2\ge0$, and 
\begin{equation}
\label{secondo010}
\Psi(m,\varepsilon,p)
 \!:=\!
 \frac{\alpha}{12}m^2(3m^2\!-8b\varepsilon m+6b^2\varepsilon^2)
 +\!
 \Psi_\rr{B}(m,\varepsilon,p)
\end{equation}
where 
\begin{equation}
\label{secondo020}
\Psi_\rr{B}(m,\varepsilon;p):=
 p\varepsilon+\frac{1}{2}\varepsilon^2+\frac{1}{2}a(m-b\varepsilon)^2
\end{equation}
is the Biot potential energy density~\cite{biot01},
$a>0$ is the ratio between the fluid and the solid rigidity, 
$b>0$ is a coupling between the fluid and the solid component, 
$p>0$ is the external pressure,
and
$\alpha>0$ is a material parameter responsible for the showing 
up of the additional equilibrium.
We remark that the condition 
$k_1k_3-k_2^2\ge0$ ensures that the second gradient 
part $K$ of the overall potential energy density is convex.
Under this assumption there exists a minimizer 
for the action functional 
\begin{displaymath}
\int_{\ell_1}^{\ell_2}\rr{d}X_\rr{s}\,
   \Phi(m',\varepsilon',m,\varepsilon) 
\end{displaymath}
on a bounded domain.
As we will see later to ensure the existence of a connection profile,
which is a Dirichelet problem on an unbounded domain,
it will be necessary to assume 
$k_1k_3-k_2^2>0$; in the limiting case 
$k_1k_3-k_2^2=0$ the existence of the connection will depend on the 
choice of the parameter $k_1$, $k_2$, and $k_3$.

\begin{figure}
\begin{picture}(200,150)
\put(-20,0)
{
\resizebox{8cm}{!}{\rotatebox{0}{\includegraphics{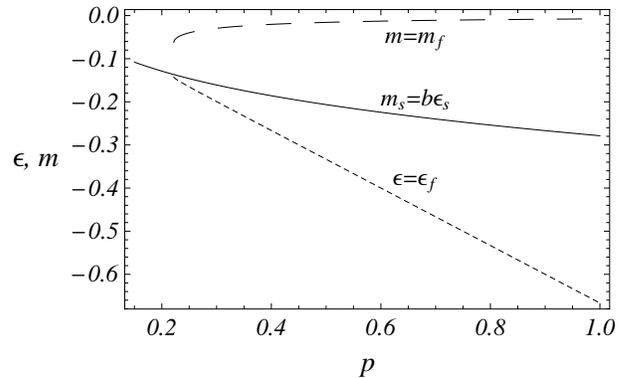}}} 
}
\end{picture}  
\caption{From the bottom to the top the graphs of $\varepsilon_\rr{f}(p)$,
$m_\rr{s}(p)=b\varepsilon_\rr{s}(p)$, and $m_\rr{f}(p)$
for $a=0.5$, $b=1$, and $\alpha=100$.
}
\label{f:fig-emme-eps} 
\end{figure}

We have already studied~\cite{CIS01} 
the associated first gradient model with 
overall potential energy density $\Psi$ and we have proven 
the existence of a phase transition driven by the 
external pressure $p$.
More precisely it has been shown that there exists a \textit{critical
pressure} $p_\rr{c}=p_\rr{c}(\alpha,a,b)$ such that 
for $0<p\le p_\rr{c}$ the system admits the single \textit{standard} phase 
$(m_\rr{s}(p),\varepsilon_\rr{s}(p))$, 
while 
a second \textit{fluid--rich} phase  
$(m_\rr{f}(p),\varepsilon_\rr{f}(p))$, 
appears
for $p>p_\rr{c}$.
The standard phase is similar to 
the unique phase described by the model 
with potential energy density\cite{biot01} $\Psi_\rr{B}$.
In FIG.~\ref{f:fig-emme-eps} the standard 
and the fluid--rich phases are depicted for $p\ge p_\rr{c}$ and for a 
particular choice of the physical parameters $\alpha,a,b$.

\section{Coexistence}
\label{s:coesistenza}
\par\noindent
The second gradient model has the same 
phases as the associated first gradient model.
The main result of this paper is the existence 
of $p_\rr{co}=p_\rr{co}(\alpha,a,b)>p_\rr{c}$, 
called \emph{coexistence pressure}, such that 
the standard and the fluid--rich phases coexist at the pressure 
$p=p_\rr{co}$ and do not coexist at $p>p_\rr{c}$ and $p\neq p_\rr{co}$.
The proof will be achived in two steps: first we shall 
show that there exist a unique value of the pressure such that 
the total potential energy densities evaluated at the two phases 
are equal; the second step will consist in proving the existence of the 
connection, that is the equilibrium profile connecting the two phases.

\subsection{Coexistence pressure}
\label{s:coepre}
\par\noindent
We first review some of the results in the previous
paper~\cite{CIS01,CIS02};
there we have studied the equations $\Psi_\varepsilon=0$ 
and $\Psi_m=0$ looking for the minima 
of the function $\Psi$.

We have shown that the standard phase 
$(m_\rr{s}(p),\varepsilon_\rr{s}(p))$ 
is the solution of the two equations 
$m=b\varepsilon$ and 
$p=f_1(\varepsilon)$, for any $p>0$, where 
$f_1(\varepsilon):=-\varepsilon-\alpha b^4\varepsilon^3/3$.

On the other hand 
the fluid--rich phase 
$(m_\rr{f}(p),\varepsilon_\rr{f}(p))$ 
is the solution, with the smallest value of $\varepsilon$,
of the two equations 
$m=m_+(\varepsilon)$ and 
$p=f_+(\varepsilon)$, where 
\begin{displaymath}
m_+(\varepsilon)=
\frac{b}{2}
\Big[
     \varepsilon+\sqrt{\varepsilon^2-\frac{4a}{\alpha b^2}}
\Big]
\end{displaymath}
and
\begin{displaymath}
f_+(\varepsilon)
\!:=\!
-\varepsilon+ab[m_+(\varepsilon)-b\varepsilon]
  -\alpha b^2\varepsilon m_+^2(\varepsilon)
  \vphantom{\bigg\{}
  +\frac{2}{3}\alpha bm_+^3(\varepsilon)
\end{displaymath}
For $\varepsilon\le-2/(b\sqrt{\alpha/a})$ 
the function $f_+(\varepsilon)$ is positive, diverging to $+\infty$ 
for $\varepsilon\to-\infty$, and has a minimum at $\varepsilon_\rr{c}$ 
such that $f_+(\varepsilon_\rr{c})=p_\rr{c}$; this explains why the fludized 
phase is seen only for $p>p_\rr{c}$.
Moreover it has been proven that for any $p>p_\rr{c}$ the point 
$(m_\rr{f}(p),\varepsilon_\rr{f}(p))$ 
is a minimum of the two variable potential 
energy $\Psi(m,\varepsilon,p)$ with $p$ fixed, while it is a saddle 
point for $p=p_\rr{c}$.

\begin{figure}
\begin{picture}(200,150)
\put(-20,0)
{
\resizebox{8cm}{!}{\rotatebox{0}{\includegraphics{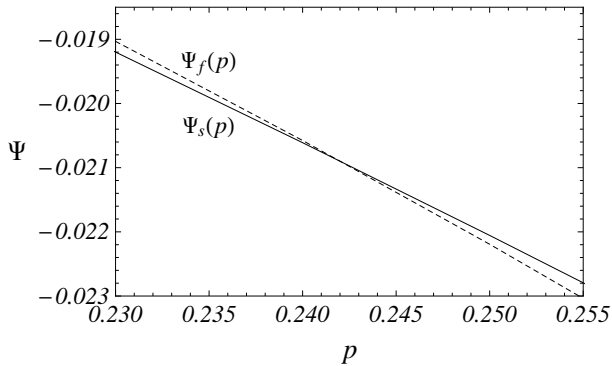}}}
}
\end{picture}
\caption{Graph of the overall potential energy $\Psi_\rr{s}(p)$ and
$\Psi_\rr{f}(p)$ for $a=0.5$, $b=1$, and $\alpha=100$.
}
\label{f:ene1}
\end{figure}

We now prove the first step of the above
stated coexistence result.
For any $p>p_\rr{c}$, we let
$\Psi_\rr{s}(p):=\Psi(m_\rr{s}(p),\varepsilon_\rr{s}(p),p)$
and 
$\Psi_\rr{f}(p):=\Psi(m_\rr{f}(p),\varepsilon_\rr{f}(p),p)$
and prove that 
\begin{equation}
\label{coes00}
\begin{array}{l}
\Psi_\rr{s}(p)
>
\Psi_\rr{f}(p)
\;\;\textrm{ for }p>p_\rr{co}
\\
\Psi_\rr{s}(p)
=
\Psi_\rr{f}(p)
\;\;\textrm{ for }p=p_\rr{co}
\\
\Psi_\rr{s}(p)
<
\Psi_\rr{f}(p)
\;\;\textrm{ for }p_\rr{co}>p\ge p_\rr{c}
\end{array}
\end{equation}
that is the overall potential energy density of the standard and the 
fluid--rich phases are equal only at the coexistence pressure.
This statement has been tested on
numerical grounds, see FIG.~\ref{f:ene1} where the graphs of the
functions $\Psi_\rr{s}(p)$ and $\Psi_\rr{f}(p)$
are depicted for a given set of
physical parameters.

In order to prove (\ref{coes00}) we first compute the derivative 
of the two functions $\Psi_\rr{s}(p)$ and $\Psi_\rr{f}(p)$ (with respect to 
$p$); by using 
(\ref{secondo010}), the chain rule, and 
the fact that  
$(m_\rr{s}(p),\varepsilon_\rr{s}(p))$
and 
$(m_\rr{f}(p),\varepsilon_\rr{f}(p))$
are solutions of the equations $\Psi_m(m,\varepsilon,p)=0$ and
$\Psi_\varepsilon(m,\varepsilon,p)=0$, we have that 
$\Psi_\rr{s}'(p)=\varepsilon_\rr{s}(p)$,
$\Psi_\rr{f}'(p)=\varepsilon_\rr{f}(p)$,
$\Psi_\rr{s}''(p)=\varepsilon_\rr{s}'(p)$,
and
$\Psi_\rr{f}''(p)=\varepsilon_\rr{f}'(p)$.

Now, 
since $\varepsilon_\rr{s}(p)$ and $\varepsilon_\rr{f}(p)$ 
are negative functions of the 
pressure, we have that both $\Psi_\rr{s}(p)$ and $\Psi_\rr{f}(p)$ are 
decreasing functions of the pressure on the interval $(p_\rr{c},+\infty)$.
Moreover, noted that both 
$f_1$ and $f_+$ are decreasing functions (of the strain) on 
$(-\infty,\varepsilon_\rr{c}]$, we have that 
$\varepsilon_\rr{s}(p)$ and $\varepsilon_\rr{f}(p)$ decrease
when $p$ increases.
It then follows that 
$\varepsilon_\rr{s}'(p)$ and $\varepsilon_\rr{f}'(p)$ are negative and
therefore
$\Psi_\rr{s}(p)$ and $\Psi_\rr{f}(p)$ are concave 
on the interval $(p_\rr{c},+\infty)$.

Since the two functions 
$\Psi_\rr{s}(p)$ and $\Psi_\rr{f}(p)$ are decreasing concave functions 
on the interval $(p_\rr{c},+\infty)$, in order to prove (\ref{coes00}) 
it is sufficient to show that
$\Psi_\rr{s}(p_\rr{c})<\Psi_\rr{f}(p_\rr{c})$ and 
$\Psi_\rr{s}(p)>\Psi_\rr{f}(p)$ for some $p$ sufficiently large.
The proof of the first remark is easy:
at $p=p_\rr{c}$ the two variable function 
$\Psi(m,\varepsilon,p_\rr{c})$ has just the 
two stationary points\cite{CIS01}
$(m_\rr{s}(p_\rr{c}),\varepsilon_\rr{s}(p_\rr{c}))$
and
$(m_\rr{f}(p_\rr{c}),\varepsilon_\rr{f}(p_\rr{c}))$.
Since $(m_\rr{s}(p_\rr{c}),\varepsilon_\rr{s}(p_\rr{c}))$ is a local 
minimum of $\Psi(m,\varepsilon,p_\rr{c})$,
which tends to $+\infty$ along every 
direction on the plane $m$--$\varepsilon$, the single local minimum 
must be the absolute minimum; hence, 
$\Psi_\rr{s}(p_\rr{c})<\Psi_\rr{f}(p_\rr{c})$.
The second remark follows from 
the asymptotic behavior of the two functions 
$\Psi_\rr{s}(p)$ and $\Psi_\rr{f}(p)$; as proven in 
the Appendix~\ref{s:asymptotic}, 
for $p\to\infty$ we have 
\begin{equation}
\label{coes50}
\Psi_\rr{s}(p)=-\frac{3}{4}\,3^{1/3}\,\Big(\frac{1}{\alpha b^4}\Big)p^{4/3}
          +O(p^{2/3})
\end{equation}
and 
\begin{equation}
\label{coes60}
\begin{array}{rcl}
\Psi_\rr{f}(p)
&=&
{\displaystyle
 -\frac{1}{1+ab^2}p^2+\frac{1}{2(1+ab^2)}p^2 +O(p)
 \vphantom{\bigg\{_\big\}}
}
\\
&=&
{\displaystyle
 -\frac{1}{2}(1+ab^2)p^2+O(p)
}
\\
\end{array}
\end{equation}
By comparing
the two asymptotic formulas (\ref{coes50}) and (\ref{coes60}) 
we get immediately that for $p$ large enough 
$\Psi_\rr{s}(p)>\Psi_\rr{f}(p)$.

\subsection{Connection profile}
\label{s:connessione}
\par\noindent
It is worth remarking that the variational problem (\ref{variazionale03}) 
for profiles with fixed values at the end points $\ell_1$ and 
$\ell_2$ of the interval $B_\rr{s}$, is nothing but the Hamilton principle 
for a two degree of freedom mechanical system with Lagrangian coordinates
$\varepsilon$ and $m$, kinetic energy $T$ and potential energy
$U$
respectively given by 
\begin{equation}
\label{secondo02}
T(m',\varepsilon')
=\frac{1}{2}[k_1(\varepsilon')^2+2k_2\varepsilon' m'+k_3(m')^2]
\end{equation}
and
\begin{equation}
\label{secondo03}
U(m,\varepsilon)
=-\Psi(m,\varepsilon)
\end{equation}
and the space variable $X_\rr{s}$ interpreted as time.
In other words the function $\Phi$ 
defined by (\ref{modello}), 
(\ref{secondo000}), and (\ref{secondo010}) is the 
Lagrangian for such a two degree of freedom equivalent mechanical system.

It is important to remark that the mechanical interpretation is 
correct only when $T$ is a positive definite quadratic form.
It is easy to prove that this is the case provided 
$k_1k_3-k_2^2>0$.
In the limiting case $k_1k_3-k_2^2=0$ the form $T$ is positive semidefinite, 
indeed if we substitute $k_1=k_2^2/k_3$ in (\ref{secondo02}) the function
$T$ becomes
\begin{equation}
\label{secondo05}
T(m',\varepsilon')
=K(m',\varepsilon')
=\frac{1}{2}k_3(k\varepsilon'+m')^2
\end{equation}
where we have set $k:=k_2/k_3$, and 
is equal to zero when $k\varepsilon'+m'=0$.

We study now the case $k_1k_3-k_2^2>0$ and 
postpone the degenerate 
$k_1k_3-k_2^2=0$ to the following section.
Let us denote $X_\rr{s}$ by $t$ and the derivative taken 
with respect to $t$ by the dot.
By using (\ref{moto06}) with $\Phi=T-U$ and recalling 
(\ref{secondo02}), we have that the equations of motion are 
\begin{equation}
\label{motoeq}
k_2\ddot m+k_1\ddot\varepsilon=-\frac{\partial U}{\partial\varepsilon}
\;\textrm{ and }\;
k_3\ddot m+k_2\ddot\varepsilon=-\frac{\partial U}{\partial m}
\end{equation}
We note that the mechanical energy of the associated 
mechanical problem  
$E(\dot m,\dot\varepsilon,m,\varepsilon)
 :=T(\dot m,\dot\varepsilon)+U(m,\varepsilon)$
is a constant of the motion.

First note that the two points 
$(m_\rr{s}(p),\varepsilon_\rr{s}(p))$
and
$(m_\rr{f}(p),\varepsilon_\rr{f}(p))$, with $p>p_\rr{c}$, are 
maxima of the potential energy $U(m,\varepsilon)$ of the equivalent
mechanical system.
The problem of the existence of a connection between the 
standard and the fluid-rich phase can be rephrased as follows:
look for a solution of the equations (\ref{motoeq}),
namely, a motion $(m_p(t),\varepsilon_p(t))$
of the equivalent mechanical system, 
on $\bb{R}$ connecting the phase space point 
$(m_p(-\infty),\varepsilon_p(-\infty))
  =(m_\rr{s}(p),\varepsilon_\rr{s}(p))$
and 
$(\dot m_p(-\infty),\dot\varepsilon_p(-\infty))=(0,0)$
to the phase space point 
$(m_p(+\infty),\varepsilon_p(+\infty))
 =(m_\rr{f}(p),\varepsilon_\rr{f}(p))$
and 
$(\dot m_p(+\infty),\dot\varepsilon_p(+\infty))=(0,0)$.
The connection we are 
seeking for is an heteroclinic solution of the equation of motion 
tending to two fixed points in the phase space 
for $t\to-\infty$ and $t\to+\infty$.

Recall that the mechanical energy $E$ is a constant of the motion 
and remark that at the equilibrium points it is equal to
$E(0,0,m_\rr{s}(p),\varepsilon_\rr{s}(p))
    =-\Psi(m_\rr{s}(p),\varepsilon_\rr{s}(p))$
and
$E(0,0,m_\rr{f}(p),\varepsilon_\rr{f}(p))
     =-\Psi(m_\rr{f}(p),\varepsilon_\rr{f}(p))$.
From the results in Section~\ref{s:coesistenza} it follows that 
those two energies are equal only for $p=p_\rr{co}$. 
This remark yields that for any $p>p_\rr{c}$ and $p\neq p_\rr{co}$ the 
standard and the fluid--rich phases do not coexist.

We are left with the case $p=p_\rr{co}$. In principle an heteroclinic 
solution can exist, but to prove its existence is an highly not trivial 
problem which has been solved, under suitable hypotheses on the potential 
energy, 
in the recent paper\cite{AF} whose main results have been 
summarized in the Appendix~\ref{s:fusco}.
Since in the not degenerate case the form $T$ is positive definite, it is 
possible to find an orthogonal transformation of the coordinates 
in the plane $m$--$\varepsilon$ which diagonalize the form itself. 
Then,
performing this transformation and subctracting to the potential 
energy $U$ of the equivalent mechanical system the constant term 
$U(m_\rr{s}(p_\rr{co}),\varepsilon_\rr{s}(p_\rr{co}))=
 U(m_\rr{f}(p_\rr{co}),\varepsilon_\rr{f}(p_\rr{co}))$, the problem 
of finding a connection between the standard and fluid--rich phase 
is transformed in a problem in the form (\ref{eak})
with $n=2$ and $W$ replaced by 
$-[U-U(m_\rr{s}(p_\rr{co}),\varepsilon_\rr{s}(p_\rr{co}))]$.
Since this function satisfies the hypotheses of the 
Theorem~3.6 by Alikakos and Fusco\cite{AF} 
(see the Appendix~\ref{s:fusco})
we can then conclude that in the case $p=p_\rr{co}$ there exists a 
connection between the standard and the fluid--rich phase and hence 
the two phases coexist.

\begin{figure}
\begin{picture}(200,180)
\put(10,0)
{
\resizebox{6cm}{!}{\rotatebox{0}{\includegraphics{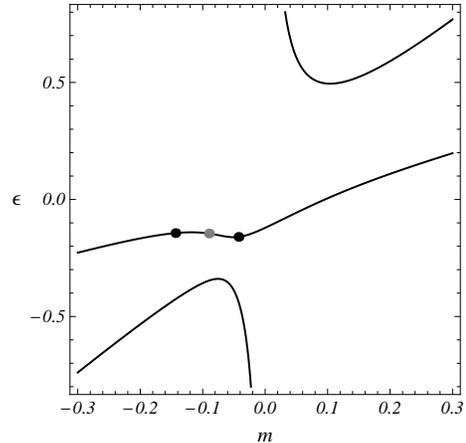}}} 
}
\end{picture}  
\caption{Graph of the constraint curve
in the plane $m$--$\varepsilon$.
The three disks represent 
the fluid--rich phase, the standard phase and the saddle (gray disk) 
of the potential energy $\Psi$. 
Parameters: $a=0.5$, $b=1$, $\alpha=100$, $k_3=1$, $k=1$,
and $p=p_\rr{co}=0.24218$.
}
\label{f:vincolo1} 
\end{figure} 

\section{The degenerate case}
\label{s:singolare}
\par\noindent
Consider the case $k_2=\pm\sqrt{k_1k_3}$ and 
the change of variables
\begin{displaymath}
x:=\frac{m+k\varepsilon}{\sqrt{1+k^2}}
\;\;\;\textrm{ and }\;\;\; 
y:=\frac{-km+\varepsilon}{\sqrt{1+k^2}}
\end{displaymath}
where we recall $k=k_2/k_3=\pm\sqrt{k_1/k_3}$,
which amounts to perform a rotation
of the cartesian reference system in the plane $m$--$\varepsilon$.
Using the new variables the two functions $T$ and $U$
become respectively
\begin{equation}
\label{secondo07}
K(\dot x,\dot y)
=
T(\dot m(\dot x,\dot y),\dot\varepsilon(\dot x,\dot y))
=
\frac{1}{2}k_3(1+k^2)\dot x^2
\end{equation}
and
\begin{equation}
\label{secondo08}
V(x,y)
=
U(m(x,y),\varepsilon(x,y))
\end{equation}
where we have used (\ref{secondo05}) and (\ref{secondo03}) with
\begin{displaymath}
m=
  \frac{x-ky}{\sqrt{1+k^2}}
\;\;\;\textrm{ and }\;\;\;
\varepsilon=
  \frac{kx+y}{\sqrt{1+k^2}}
\end{displaymath}
The expression (\ref{secondo08}) of $V$ is awful, but 
this will not be a problem since $V$ is precisely the two variable functions
$-\Psi$, which we have already deeply studied\cite{CIS01}, 
written via a rotation of the cartesian reference system.

We apply, now, the variational principle (\ref{variazionale03}) to the 
total poroelastic potential energy density
$\mathscr{L}(\dot x,\dot y,x,y):=K(\dot x,\dot y)-V(x,y)$ and get 
the analogous (indeed it is a particularization) of the equations 
(\ref{moto06})
\begin{equation}
\label{secondo10}
\frac{\partial\mathscr{L}}{\partial x}-
\frac{\rr{d}}{\rr{d}t}\frac{\partial\mathscr{L}}{\partial\dot x}=0
\;\;\;\textrm{ and }\;\;\;
\frac{\partial\mathscr{L}}{\partial y}=0
\end{equation}
which must be solved with the boundary conditions (\ref{moto07}). 
By using the definition of $\mathscr{L}$ the above equations 
become
\begin{equation}
\label{secondo12}
k_3(1+k^2)
\ddot x=-\frac{\partial V}{\partial x}(x,y)
\;\;\;\textrm{ and }\;\;\;
\frac{\partial V}{\partial y}(x,y)=0
\end{equation}
We remark that the second of the equations above is an algebraic 
equation involving the two variable $x$ and $y$; provided it can be 
solved w.r.t.\ $y$, the first one becomes a 
second order ordinary differential equation in the unique 
unknown function $x$. 
More precisely, the root locus of $\partial V(x,y)/\partial y=0$ is made of 
a certain number of maximal components 
such that each of them is the graph of a 
function $x\in\bb{R}\to y(x)\in\bb{R}$;
for each of them the first of the two equations
(\ref{secondo12}) becomes
a standard one dimensional conservative mechanical system 
with potential energy $V(x,y(x))$.

\begin{figure}
\begin{picture}(200,180)
\put(10,0)
{
\resizebox{6cm}{!}{\rotatebox{0}{\includegraphics{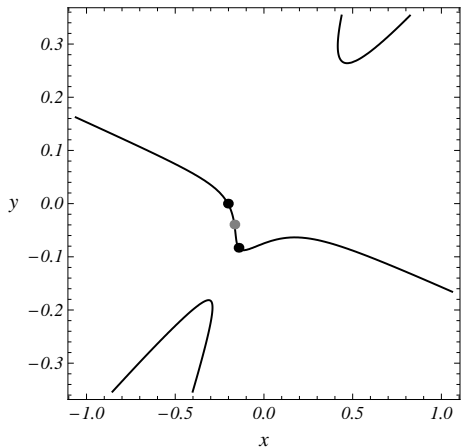}}} 
}
\end{picture}  
\caption{Graph of the constraint curve
in the plane $x$--$y$.
The three disks represent 
the fluid--rich phase, the standard phase and the saddle (gray disk) of the 
potential energy $\Psi$. 
Parameters: $a=0.5$, $b=1$, $\alpha=100$, $k_3=1$, $k=1$,
and $p=p_\rr{co}=0.24218$.
}
\label{f:vincolo1xy} 
\end{figure} 

\subsection{The degenerate case: heteroclinic}
\label{s:deghet}
\par\noindent
The function $V$ is obtained by flipping the sign of the 
function $\Psi$ and rotating the coordinate axes.
This implies that the function 
$V$, at $p=p_\rr{co}$, has the two absolute maximum points 
\begin{displaymath}
(x_\rr{s}(p),y_\rr{s}(p))
 =\Big(\frac{m_\rr{s}(p)+k\varepsilon_\rr{s}(p)}{\sqrt{1+k^2}},
       \frac{-km_\rr{s}(p)+\varepsilon_\rr{s}(p)}{\sqrt{1+k^2}}\Big)
\end{displaymath}
and 
\begin{displaymath}
(x_\rr{f}(p),y_\rr{f}(p))
 =\Big(\frac{m_\rr{f}(p)+k\varepsilon_\rr{f}(p)}{\sqrt{1+k^2}},
       \frac{-km_\rr{f}(p)+\varepsilon_\rr{f}(p)}{\sqrt{1+k^2}}
  \Big)
\end{displaymath}
corresponding, respectively, to the standard and to the fluid--rich phases. 

Since $(m_\rr{s}(p),\varepsilon_\rr{s}(p))$
and $(m_\rr{f}(p),\varepsilon_\rr{f}(p))$
satisfy the equations $\Psi_m(m,\varepsilon)=0$ and 
$\Psi_\varepsilon(m,\varepsilon)=0$, we have that the two points
$(x_\rr{s}(p),y_\rr{s}(p))$ and 
$(x_\rr{f}(p),y_\rr{f}(p))$ are solutions of the constraint equation 
$\partial V(x,y)/\partial y=0$
and hence they belong to the constraint curve. 

\begin{figure}
\begin{picture}(200,180)
\put(-20,0)
{
\resizebox{8cm}{!}{\rotatebox{0}{\includegraphics{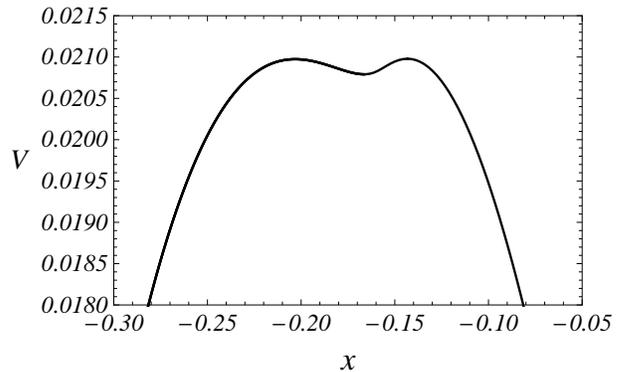}}} 
}
\end{picture}  
\caption{Function $V(x,y(x))$ in the case
$a=0.5$, $b=1$, $\alpha=100$, $k_3=1$, $k=1$,
and $p=p_\rr{co}=0.24218$.
}
\label{f:vdix} 
\end{figure} 

We consider, now, the case in which 
at $p=p_\rr{co}$ the two points above 
fall on the same maximal component of the constraint equation
(see FIG.~\ref{f:vincolo1} and \ref{f:vincolo1xy}). 
Using the conservation of the mechanical energy of the equivalent 
one dimensional conservative system allows for reducing
the computation of the coexistence profile (heteroclinic) 
to the evaluation of a definite integral. 
Since the function $V$ has two isolated absolute
maximum points which, by hypothesis, belong to the 
same maximal component of the 
constraint curve, we have that the function $V(x,y(x))$ of the 
real function $x$ has two absolute isolated maxima in 
$x_\rr{s}(p_\rr{co})$ and $x_\rr{f}(p_\rr{co})$ 
(see FIG.~\ref{f:vdix}).
Consider the motion of the equivalent one dimensional 
system corresponding to the energy level
$V_{\rr{max}}:=V(x_\rr{s}(p_\rr{co}),y_\rr{s}(p_\rr{co}))$.
The conservation of the mechanical energy
implies 
\begin{displaymath}
\frac{1}{2}k_3(1+k^2)\dot{x}^2+V(x,y(x))
=
V_{\rr{max}}
\end{displaymath}
Hence, the heteroclinic connecting the two 
maxima is given by
\begin{equation}
\label{etero}
t=\int_{x_0}^x
    \sqrt{
    \frac{k_3(1+k^2)}
         {2[V_{\rr{max}}-V(x',y(x'))]}
         }
    \,\rr{d}x'
\end{equation}
for any 
$x\in(\min\{x_\rr{s}(p_\rr{co}),x_\rr{f}(p_\rr{co})\}
          ,\max\{x_\rr{s}(p_\rr{co}),x_\rr{f}(p_\rr{co})\})$
and for some fixed 
$x_0$ in the same interval.
By changing $x_0$ it is found a family of heteroclinic orbits 
which are the same curve up to a time translation.

\begin{figure*}
\begin{picture}(200,250)
\put(-100,0)
{
\resizebox{14cm}{!}{\rotatebox{0}{\includegraphics{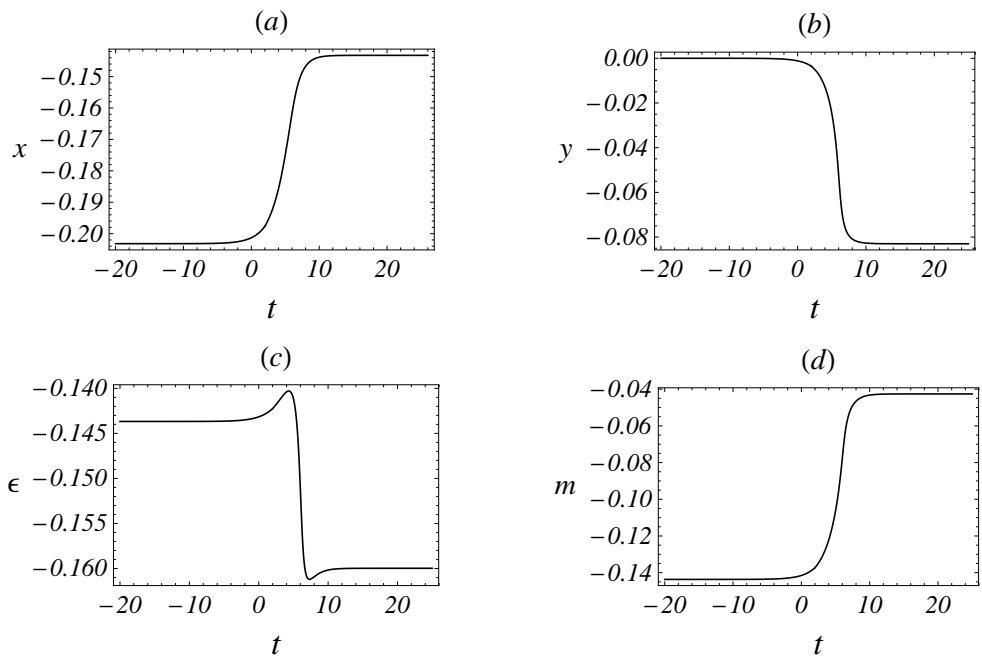}}} 
}
\end{picture}  
\caption{Heteroclinic (coexistence profile) in the case
$a=0.5$, $b=1$, $\alpha=100$, $k_3=1$, $k=1$, and $p=p_\rr{co}=0.24218$.
Time (space in the original model) on the horizontal axis and 
$x$, $y$, $\varepsilon$, and $m$ 
on the vertical axis respectively in (a), (b), (c), and (d).
}
\label{f:etero} 
\end{figure*} 

\begin{figure}
\begin{picture}(200,180)
\put(-20,0)
{
\resizebox{8cm}{!}{\rotatebox{0}{\includegraphics{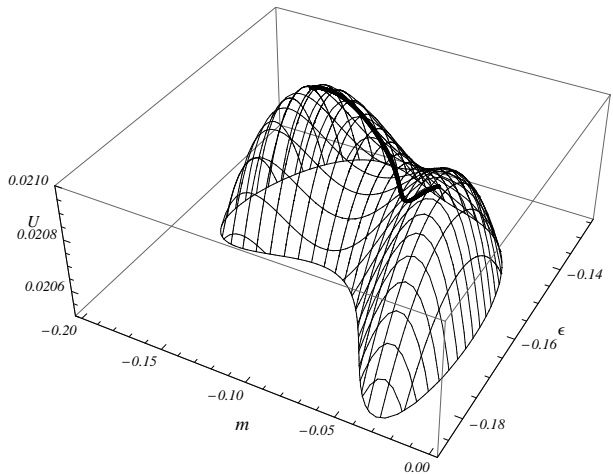}}} 
}
\end{picture}  
\caption{Graph of the 
function $U(m,\varepsilon)$ and the heteroclinic connecting 
the two maxima.
Parameters: $a=0.5$, $b=1$, $\alpha=100$, $k_3=1$, $k=1$,
and $p=p_\rr{co}=0.24218$.
}
\label{f:etero2} 
\end{figure} 

Results are depicted in the FIG.~\ref{f:etero}.
The $\varepsilon$ coexistence profile shows a bump\cite{dawson,lgy}
close both to the 
standard and the fluid--rich phase. This behavior is due to the 
two--dimensionality of the problem: in FIG.~\ref{f:etero2}
we have depicted the heteroclinic on the graph of the function 
$U(m,\varepsilon)=-\Psi(m,\varepsilon)$. From the picture it is clear 
that the optimal path climbs the two hills going around the hills
themselves.
In other words the existence of the bump in the connecting 
$\varepsilon$--profile is due to 
the shape of the constraint curve in the plane $m$--$\varepsilon$.
Since the problem has been reduced to the computation of the 
heteroclinic of a one dimensional conservative 
mechanical system in the $x$ variable, it is obvious that no 
bump can exist in the $x$--profile. On the other hand 
by looking closely at the picture in FIG.~\ref{f:vincolo1} and 
\ref{f:vincolo1xy}, it emerges that the constraint curve is monotonic 
w.r.t.\ $y$ and $m$; this implies the monotonicity of the 
$y$ and $m$--profiles. 
However, 
it is possible to find values of the parameters such that 
the $y$--profile presents a bump.

\begin{figure}
\begin{picture}(200,180)
\put(10,0)
{
\resizebox{6cm}{!}{\rotatebox{0}{\includegraphics{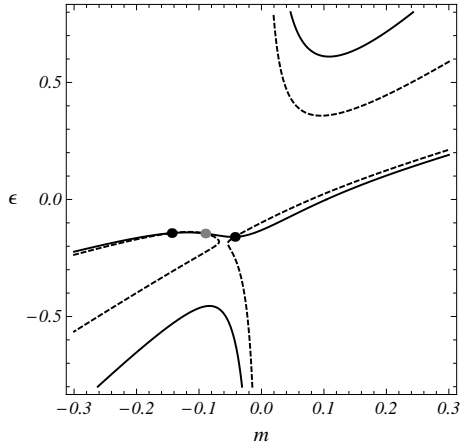}}} 
}
\end{picture}  
\caption{The constraint curves and the points representing 
the fluid--rich phase, the standard phase and the saddle (gray disk) of the 
potential energy $\Psi$. 
Parameters: $a=0.5$, $b=1$, $\alpha=100$, $k_3=1$, 
$k=0.7,1.9$ (solid, dashed),
and $p=p_\rr{co}=0.24218$.
}
\label{f:kpos} 
\end{figure}

\begin{figure}
\begin{picture}(200,180)
\put(10,0)
{
\resizebox{6cm}{!}{\rotatebox{0}{\includegraphics{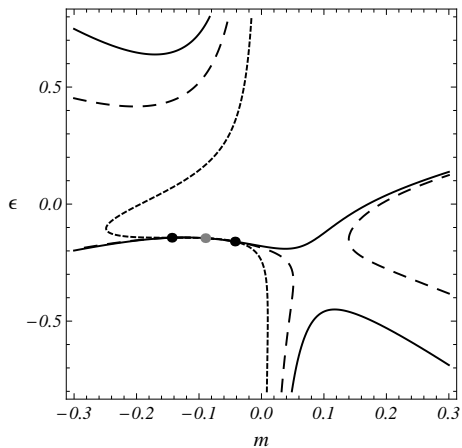}}} 
}
\end{picture}  
\caption{The constraint curves and the points representing 
the fluid--rich phase, the standard phase and the saddle (gray disk) of the 
potential energy $\Psi$. 
Parameters: $a=0.5$, $b=1$, $\alpha=100$, $k_3=1$, 
$k=-0.3,-0.4,-1.0$ (solid, dashed, dotted),
and $p=p_\rr{co}=0.24218$.
}
\label{f:kneg} 
\end{figure}

In Section~\ref{s:deghet} we have proven that in the not degenerate
case the connecting profile does exist for any proper choice of the 
parameters. A similar 
result does not hold true in the degenerate case, 
indeed it is possible to 
find the connection if and only if
the two maxima of the function $U$ lie on the same maximal 
component of the constraint curve. 
We have that this is not the case for $k>0$ large enough,
see the dashed curve in 
FIG.~\ref{f:kpos} which is associated to the value
$k=1.9$. It is immediate 
to remark that the two maxima do not lie on the same connected 
component, hence in this case it is not possible to find a connection 
between the fluid--rich and the standard phase.
It is worth remarking that no evidence of this patologic behavior 
is found in the case $k<0$; 
see FIG.~\ref{f:kneg} where the constraint curve and the stationary 
point of $U$ are depicted for 
$a=0.5$, $b=1$, $\alpha=100$, 
$k_3=1$, $k=-0.3,-0.4,-1.0$, and $p=p_\rr{co}=0.24218$. 
This case is the most
interesting one from the physical point of view, indeed for $k_2<0$ 
the coupling between $\varepsilon'$ and $m'$ is negative, hence 
the preferred states are such that the two fields $\varepsilon(X_\rr{s})$ 
and $m(X_\rr{s})$ are both increasing or decreasing.

\subsection{The degenerate case: homoclinic}
\label{s:deghom}
\par\noindent
Consider the degenerate model and suppose that the pressure $p$ is 
larger than $p_\rr{c}$ but different from $p_\rr{co}$. 
Suppose that the two local minima of the potential energy $\Psi$ 
lie on the same connected component of the constraint curve whose 
equation is (\ref{secondo12}).
Consider the function $V(x,y(x))$ as in Section~\ref{s:deghet} and note that 
the two local maxima are not equal. 

\begin{figure*}[t]
\begin{picture}(200,250)
\put(-100,0)
{
\resizebox{14cm}{!}{\rotatebox{0}{\includegraphics{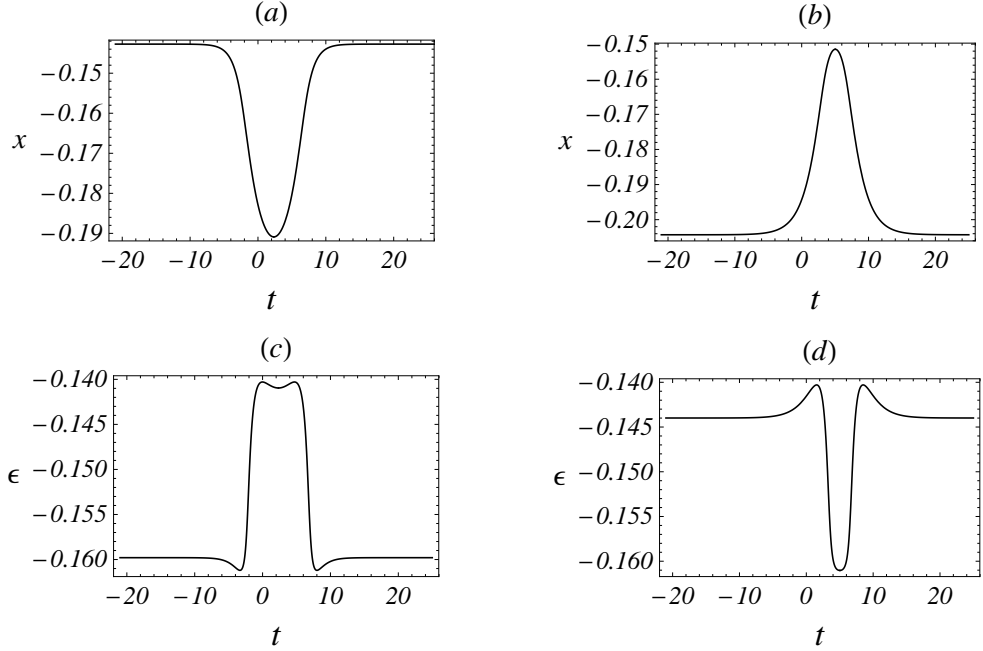}}} 
}
\end{picture}  
\caption{Homoclinic profile (critical nucleus) in the case
$a=0.5$, $b=1$, $\alpha=100$, $k_3=1$, $k=1$, and 
$p=p_\rr{co}-10^{-2}p_\rr{co}$ (left) and
$p=p_\rr{co}+10^{-2}p_\rr{co}$ (right) with 
$p_\rr{co}=0.24218$.
Time (space in the original model) on the horizontal axis and 
$x$ ((a) and (b)) and $\varepsilon$ ((c) and (d))
on the vertical one.
}
\label{f:omo} 
\end{figure*}

We consider the homoclinic solution corresponding to the lowest maximum
$\bar x$.
In analogy with the discussion of the above section, 
the homoclinic equilibrium profile can be found, see
FIG.~\ref{f:omo}, by computing the integral 
\begin{equation}
\label{omoclinica}
t=\pm\int_{\hat x}^x
    \sqrt{
          \frac{k_3(1+k^2)}
               {2[V(\bar x,y(\bar x))-V(x',y(x'))]} 
         }
    \,\rr{d}x'
\end{equation}
for any 
$x\in(\min\{\hat x,\bar x\},
      \max\{\hat x,\bar x\})$
with 
$\hat x$ the unique (inversion) point in 
the interval 
$(\min\{x_\rr{st},x_\rr{f}\},\max\{x_\rr{st},x_\rr{f}\})$
such that 
$V(\bar x,y(\bar x))=V(\hat x,y(\hat x))$.

The homoclinic solution corresponding to the lowest maximum
is often interpreted 
as a ``critical nucleus.'' 
In the sense that, if a dynamic evolution would be 
taken into account, one would expect that an initial condition 
close to the critical nucleus would evolve into the standard or the 
fluid--rich phase (subcritical and supercritical behavior). 
This behavior depends on the size of the 
droplet in the neighborhood of $t=0$. 
Indeed in $t=0$ the profile has the value $\hat x$ which, for 
$p$ close to $p_\rr{co}$, is a good approximation of the 
phase corresponding to the largest maximum of the function $V$.
In other words the critical nucleus can be seen as a droplet 
of the phase correpondig to the smallest value of the 
potential energy $\Psi$ plunged into the other phase. 

\section{Conclusions}
\label{s:conclusioni}
\par\noindent
In conclusion we have studied the phase transition 
between the fluid poor and rich phases
in the context of consolidating completely fluid saturated porous media. 
A second gradient model to study the existence 
of such a transition has been proposed.
Moreover, coexistence between the two phases at the 
pressure $p_\rr{co}$, defined as the pressure such that 
the total potential energy of the two phases is the same,
has been established.
We have also shown that at different values of the pressure the 
two phases cannot coexist. For a particular choice of the 
parameters of the model it is possible 
to reduce the problem of finding the coexistence profile to 
the computation of a definite integral. We have studied 
the coexistence profile for different values of the physical 
parameters of the model and shown that non--monotonic interfaces 
exist. 

\begin{acknowledgements}
The authors are grateful to G.\ Fusco for helpful discussions and hints.
\end{acknowledgements}

\appendix

\section{Asymptotic behavior of potential energy}
\label{s:asymptotic}
\par\noindent
In this appendix 
we discuss the asymptotic behavior of the two functions 
$\Psi_\rr{s}(p)$ and $\Psi_\rr{f}(p)$, 
see Section~\ref{s:coepre}, for $p$ large and, in particular, 
prove the equations (\ref{coes50}) and  (\ref{coes60}).
We first note that by using $m_\rr{s}(p)=b\varepsilon_\rr{s}(p)$, 
we get 
\begin{displaymath}
\Psi_\rr{s}(p)=p\varepsilon_\rr{s}(p)
          +\frac{1}{2}(\varepsilon_\rr{s}(p))^2
          +\frac{1}{12}\alpha b^4(\varepsilon_\rr{s}(p))^4
\end{displaymath}
The equation $f_1(\varepsilon)=p$ is a cubic equation
in the form $\varepsilon^3+\lambda\varepsilon+\lambda p=0$, with 
$\lambda=3/(\alpha b^4)$; by Cardano's formula, since
$D:=(\lambda/3)^3+(\lambda p/2)^2>0$, there exists a single 
real solution given by
\begin{displaymath}
\begin{array}{rrl}
\varepsilon_\rr{s}(p)
&\!=&
 \!
{\displaystyle
 \Big[-\frac{1}{2}\lambda p+\sqrt{D}\Big]^{1/3}
 +
 \Big[-\frac{1}{2}\lambda p-\sqrt{D}\Big]^{1/3}
 \vphantom{\bigg\{_\big\}}
}
\\
&\!=&
 \!
{\displaystyle
 \Big[-\frac{1}{2}\frac{3}{\alpha b^4} p
      +\sqrt{\Big(\frac{1}{3}\frac{3}{\alpha b^4}\Big)^3
             +\Big(\frac{1}{2}\frac{3}{\alpha b^4} p\Big)^2}\Big]^{1/3}
 \vphantom{\bigg\{_\big\}}
}
\\
&&
 \!
{\displaystyle
 +
 \Big[-\frac{1}{2}\frac{3}{\alpha b^4} p
      -\sqrt{\Big(\frac{1}{3}\frac{3}{\alpha b^4}\Big)^3
             +\Big(\frac{1}{2}\frac{3}{\alpha b^4} p\Big)^2}\Big]^{1/3}
}
\\
\end{array}
\end{displaymath}
By using the Taylor series 
$(1+x)^\alpha=\sum_{n=0}^\infty C_n(\alpha)x^n$, with
$C_n(\alpha)=\alpha(\alpha -1)\cdots(\alpha-n+1)/n!$ being the binomial 
coefficient, which is convergent for $-1<x<+1$, it is not difficult to prove 
that 
$\varepsilon_\rr{s}(p)=-(3/(\alpha b^4))^{1/3}p^{1/3}
           +(3/\alpha b^4)^{2/3}(1/p)^{1/3}/3+O(p^{-5/3})$
for $p$ large.
By inserting this expression in the expansion for $\Psi_\rr{s}$,
we get 
equation (\ref{coes50}).

We can perform a similar computation for $\Psi_\rr{f}(p)$.
Accounting in particular for the qualitative of $\varepsilon_\rr{f}(p)$,
which tends to $-\infty$ when $p$ is increased,
we shall study the asymptotic behavior of $m_+(\varepsilon)$ for 
$\varepsilon\to-\infty$ and that of $\varepsilon_\rr{f}(p)$ which is 
the solution of the equation $f_+(\varepsilon)=p$ when 
$p\to\infty$.
The result of this analysis will provide us with the asymptotic behavior of 
$\Psi_\rr{f}(p)$.
First of all we note that 
\begin{displaymath}
m_+(\varepsilon)=\frac{1}{2}b\varepsilon
\Big[
     \frac{2a}{\alpha b^2\varepsilon^2}
     +\frac{2a^2}{\alpha^2 b^4\varepsilon^4}
     +O(\varepsilon^{-6})
\Big]
\end{displaymath}
for $\varepsilon\to-\infty$.
By using (\ref{secondo010}) we then have
\begin{displaymath}
\Psi_\rr{f}(p)=p\varepsilon_\rr{f}(p)
          +\frac{1}{2}(1+ab^2)(\varepsilon_\rr{f}(p))^2
          -\frac{a^2}{2\alpha}
          +O\Big((\varepsilon_\rr{f}(p))^{-2}\Big)
\end{displaymath}
for $p\to\infty$,
where we have used that $\varepsilon_\rr{f}(p)\to-\infty$ for $p\to\infty$.
The function $\varepsilon_\rr{f}(p)$ is implicitly defined by the 
equation $f_+(\varepsilon)=p$ which is pretty complicated.
By expanding $f_+$ for $\varepsilon\to-\infty$
the equation becomes
$-\varepsilon(1+ab^2)+h(\varepsilon)=p$ with $h(\varepsilon)$ a function 
having limit $0$ for $\varepsilon\to-\infty$. Suppose $p$ 
is large enough and let $\varepsilon_\rr{f}(p)$ 
be the solution of the equation above; by the qualitative study
we get that $\varepsilon_\rr{f}(p)\to-\infty$ for $p\to\infty$. 
It is then easy to show that 
$g(p):=\varepsilon_\rr{f}(p)-[-p/(1+ab^2)]$ tends to zero  
for $p\to\infty$, indeed, since $\varepsilon_\rr{f}$ is the solution of the 
equation above, we have that 
\begin{displaymath}
g(p)
=\frac{(1+ab^2)\varepsilon_\rr{f}(p)+p}{1+ab^2}
=\frac{h(\varepsilon_\rr{f}(p))}{1+ab^2}\to0
\end{displaymath}
for $p\to\infty$,
where we have used that $h(\varepsilon)\to0$ for $\varepsilon\to-\infty$ and
$\varepsilon_\rr{f}(p)\to-\infty$ for $p\to\infty$.
By inserting the obtained expression of $\varepsilon_\rr{f}(p)$ in the 
above expansion of $\Psi_\rr{f}(p)$ we get 
equation (\ref{coes60}).

\section{General result on the existence of connections}
\label{s:fusco}
\par\noindent
In this appendix we briefly review the main results 
by Alikakos and Fusco\cite{AF} on the existence of connections.
Let 
$W:\bb{R}^n\to\bb{R}$, with $n\ge1$,
be a $C^2(\bb{R}^n)$ positive function
satisfying the following hypotheses:
(1) $W$ has two distinct local minima
$a_-,a_+\in\bb{R}^n$ such that 
$W(a_-)=W(a_+)=0$,
(2) $W(u)>0$ for any $u\neq a_-,a_+$,
(3) $\liminf_{|u|\to\infty} W(u)>0$,
(4) there exists $r_0$ in the open 
interval $(0,|a_--a_+|)$ such that 
for any $\xi\in\bb{R}^n$ such that $|\xi|=1$
the two maps 
$r\mapsto W(a_\pm+r\xi)$
have a strictly positive derivative for every $r\in(0,r_0)$.
Conditions (1) -- (3)
are quite natural and physically obvious; condition (4) is a 
mild technical requirement allowing 
for potential energies with $C^\infty$ contact at zeroes. 

Consider the ordinary differential equation problem
\begin{equation}
\label{eak}
\left\{
\begin{array}{l}
u_{xx}=\nabla W(u)
\\
u(-\infty)=a_-
\;\;\textrm{ and }\;\;
u(+\infty)=a_+
\end{array}
\right.
\end{equation}
where $u:\bb{R}\to\bb{R}^n$.
Solutions to the problem (\ref{eak}) are known in the literature as
\textit{heteroclinc} motions of the mechanical system or 
\textit{connection} solutions in the context of phase transitions.
The Theorem~3.6 by Alikakos and Fusco\cite{AF} states that, 
under the hypotheses discussed above, 
the problem (\ref{eak}) admits a solution.
In other words the theorem states the existence of a connection 
under very general and mild requirements on the potential 
$W$.
The proof of the theorem is based on a direct 
variational computation. More precisely the authors prove the 
existence of a critical profile  
of the action functional 
\begin{displaymath}
A(u):=
\int_{-\infty}^{\infty}
\left[ 
      \dfrac{1}{2}\vert\dot u(x)\vert^2+W(u(x))
\right]\,\rr{d}x
\end{displaymath}
on the Sobolev space 
$W^{1,2}_\rr{loc}(\bb{R},\bb{R}^n)$ of functions
$u:\bb{R}\to\bb{R}^n$ such that $u$ and its weak derivative
are in $L^2(\Omega,\bb{R}^n)$ for 
any bounded subsets $\Omega\subset\bb{R}$.
Such a critical profile is the solution of the ordinary 
differential equation problem (\ref{eak}).

Compared to the standard variational calculus,
see for instance the paragraph~8.2 in Evan's classical book\cite{saravero},
the authors have to face the lack of compactess due to the 
infinite domain $\bb{R}$ on which the solution of the 
variational problem is defined. 
This problem is overcame by using suitable costraints that are 
successivley removed. 
It is also worth noting that in the Theorem~3.7 the authors state 
that the connection is a minimizer of the action functional 
$A(u)$.


\end{document}